\documentclass[reprint, amsmath, amssymb, aps, prl, superscriptaddress, nofootinbib, nobibnotes,longbibliography]{revtex4-1}

\usepackage{graphicx}

\usepackage[utf8]{inputenc}
\usepackage{amsmath}
\usepackage{amsfonts}
\usepackage{amssymb}
\usepackage{multirow}
\usepackage[colorlinks=true, linkcolor=red, citecolor=blue]{hyperref}

\usepackage[capitalise]{cleveref}
\crefname{figure}{Fig.}{Figs.}
\Crefname{figure}{Fig.}{Figs.}

\def\({\left(}
\def\){\right)}
\def\[{\left[}
\def\]{\right]}

\usepackage{acro}
\DeclareAcronym{BH}{
	short = BH ,
	long  = black hole
}
\DeclareAcronym{PBH}{
	short = PBH ,
	long  = primordial black hole
}
\DeclareAcronym{SGWB}{
	short = SGWB ,
	long  = stochastic gravitational-wave background
}
\DeclareAcronym{BBH}{
	short = BBH ,
	long  = binary black hole
}
\DeclareAcronym{GW}{
	short = GW ,
	long  = gravitational wave
}
\DeclareAcronym{CMB}{
	short = CMB ,
	long  = cosmic microwave background
} 
\DeclareAcronym{SFR}{
	short = SFR ,
	long  = star formation rate
}

\begin{document}

\title{Constraints on the Primordial Black Hole Abundance from the First Advanced LIGO Observation Run Using the Stochastic Gravitational-Wave Background}

\author{Sai Wang}
\email{wangsai@itp.ac.cn}
\affiliation{Department of Physics, The Chinese University of Hong Kong, Shatin, N.T., Hong Kong}
\author{Yi-Fan Wang}
\email{yfwang@phy.cuhk.edu.hk}
\affiliation{Department of Physics, The Chinese University of Hong Kong, Shatin, N.T., Hong Kong}
\author{Qing-Guo Huang}
\email{huangqg@itp.ac.cn}
\affiliation{CAS Key Laboratory of Theoretical Physics, Institute of Theoretical Physics, Chinese Academy of Sciences, Beijing 100190, China}
\affiliation{School of Physical Sciences, University of Chinese Academy of Sciences, No. 19A Yuquan Road, Beijing 100049, China}
\affiliation{Synergetic Innovation Center for Quantum Effects and Applications, Hunan Normal University, Changsha 410081, China}
\author{Tjonnie G. F. Li}
\email{tgfli@phy.cuhk.edu.hk}
\affiliation{Department of Physics, The Chinese University of Hong Kong, Shatin, N.T., Hong Kong}

\begin{abstract}
	Advanced LIGO's discovery of gravitational-wave events is stimulating extensive studies on the origin of binary black holes.
	Assuming that the gravitational-wave events can be explained by binary \acl{PBH} mergers,  we utilize the upper limits on the stochastic gravitational-wave background given by Advanced LIGO as a new observational window to independently constrain the abundance of \aclp{PBH} in dark matter. 
	We show that Advanced LIGO's first observation run gives the best constraint on the \acl{PBH} abundance in the mass range $1 M_\odot \lesssim M_\text{PBH}\lesssim 100 M_\odot$, pushing the previous microlensing and dwarf galaxy dynamics constraints tighter by 1 order of magnitude. 
	Moreover, we discuss the possibility to detect the \acl{SGWB} from \aclp{PBH}, in particular from subsolar mass \aclp{PBH}, by Advanced LIGO in the near future.
\end{abstract}

\maketitle

\acresetall

\textit{Introduction.}---During the first Advanced LIGO observing run, two \ac{GW} events, GW150914 and GW151226, were observed \cite{Abbott:2016blz,Abbott:2016nmj}.
Both \ac{GW} signals are found to be consistent with the mergers of \acp{BH}.
GW150914 originated from two relatively heavy coalescing \acp{BH} with masses of $36^{+5}_{-4}  M_{\odot}$ and $29^{+4}_{-4} M_{\odot}$ \cite{Abbott:2016blz}, while GW151226 originated from two coalescing \acp{BH} with masses of $14^{+8}_{-4} M_{\odot}$ and $7^{+2}_{-2} M_{\odot}$ \cite{Abbott:2016nmj}.
The local merger rate of the \ac{BBH} mergers has been inferred to be $3.4^{+8.8}_{-2.8} ~\textrm{Gpc}^{-3}\textrm{yr}^{-1}$ for GW150914, and $36^{+95}_{-30} ~\textrm{Gpc}^{-3}\textrm{yr}^{-1}$ for GW151226 \cite{TheLIGOScientific:2016pea},
where the uncertainties are given at a $90\%$ confidence level.
These discoveries robustly demonstrate that \acp{BBH} indeed exist and can merge within the age of the Universe.

The origin of these \acp{BH} and the formation mechanism of a \ac{BBH} are still under debate.
Besides an astrophysical origin \cite{TheLIGOScientific:2016htt,Belczynski:2016obo,Belczynski:2010tb,Miller:2016krr}, the possibility that these \acp{BH} are of a primordial origin and constitute a fraction of dark matter is also considered \cite{Bird:2016dcv,Clesse:2016vqa,Sasaki:2016jop,Chen:2016pud,Kashlinsky:2016sdv,Bartolo:2016ami,Cholis:2016xvo}.
The \ac{PBH} abundance in dark matter has been constrained from a variety of observations, including microlensing events caused by massive astrophysical compact halo objects \cite{Novati:2013fxa,Mediavilla:2009um,Green:2016xgy,Axelrod:2016nkp}, the gas accretion effect of \acp{PBH} on \ac{CMB} \cite{Ali-Haimoud:2016mbv} and the nondetection of a third-order Shapiro time delay using a pulsar timing array \cite{Schutz:2016khr} (see Ref.~\cite{Carr:2016drx} and references therein). 
Nevertheless, a primordial origin of GW150914 and GW151226 has not been ruled out.

Currently, the nature of dark matter is still uncertain.
There is no definitive evidence for weakly interacting massive particles (WIMPs), a prime candidate for dark matter, from experiments such as the Particle and Astrophysical Xenon Detector (PandaX-II) \cite{Tan:2016zwf}, the Large Underground Xenon dark matter experiment (LUX) \cite{Akerib:2015rjg}, the Large Hadron Collider (LHC) \cite{ATLAS:2016hao}, the Alpha Magnetic Spectrometer (AMS-02) \cite{Accardo:2014lma} and the \textit{Fermi} Large Area Telescope (\textit{Fermi} LAT) \cite{FermiLAT:2011ab}.
The situation motivates one to consider dark matter candidates other than WIMPs such as superWIMPs, light gravitinos, hidden dark matter, sterile neutrinos and axions \cite{Feng:2010gw}.
Amongst these alternatives, \acp{PBH} are also possible candidates of dark matter \cite{Carr:2016drx}.

\Acp{PBH} could be produced by direct gravitational collapse of a primordial overdensity in the early Universe, deep in the radiation dominated era \cite{Hawking:1971ei,Carr:1974nx,GarciaBellido:1996qt,Clesse:2015wea,Dolgov:2013lba}.
At the formation redshift $z_f$, the \ac{PBH} mass is roughly equal to the horizon mass, namely $M_{\textrm{BH}}\simeq\frac{4\pi}{3}{\rho_{f}}(H^{-1}_{f})^{3}\sim30 M_{\odot}[4\times10^{11}/(1+z_{f})]^{2}$ \cite{Sasaki:2016jop}.
Different mechanisms have been proposed to form binary systems from these \acp{PBH}.
Two \acp{PBH} might pass by each other accidentally and then form a binary due to energy loss by gravitational radiation \cite{Bird:2016dcv,Clesse:2016vqa}.
To account for the estimated \ac{GW} event rate, \acp{PBH} need to contribute to most of the dark matter in this model.
On the other hand, two nearby \acp{PBH} can form a binary due to the tidal force from the third neighboring \ac{PBH} \cite{Sasaki:2016jop,Nakamura:1997sm}.
The \ac{PBH} fraction of dark matter in this model can be smaller than that of Refs.~\cite{Bird:2016dcv,Clesse:2016vqa} and still be compatible with the estimated local merger rate from the gravitational-wave detections.
The expected local merger rate of binary \ac{PBH} mergers for both these models is consistent with Advanced LIGO's estimate \cite{Bird:2016dcv,Clesse:2016vqa,Sasaki:2016jop}.
Therefore, the binary \ac{PBH} scenario is capable of explaining GW150914 and GW151226.

The \ac{SGWB} from \acp{BBH} is produced from the incoherent superposition of all the merging binaries in the Universe \cite{Regimbau:2011rp,Zhu:2011bd,Wu:2011ac,Zhu:2012xw,Wu:2013xfa,Marassi:2011si,Rosado:2011kv,TheLIGOScientific:2016wyq,TheLIGOScientific:2016dpb}.
This background is potentially measurable at Advanced LIGO's projected final sensitivity \cite{TheLIGOScientific:2016wyq}.
Recently, the \ac{SGWB} following the \ac{PBH} binary formation mechanism in Refs.~\cite{Bird:2016dcv,Clesse:2016vqa} was shown to be difficult to detect by Advanced LIGO detectors given a single-mass spectrum \cite{Mandic:2016lcn}.
The amplitude of the \ac{SGWB} from \acp{PBH} could be enhanced if \acp{PBH} cluster in subhalos and have a broad mass distribution with the width of the mass distribution $\Delta M\gtrsim10^2 M_\odot$ \cite{Clesse:2016ajp}.

In this work, we utilize the upper limit on the \ac{SGWB} given by Advanced LIGO as a new observational window to independently constrain the abundance of \acp{PBH} in dark matter, and compare it to a variety of other constraining methods mentioned above.
Moreover, we consider the \ac{SGWB} spectra from different PBH masses, particularly from subsolar mass \acp{PBH}, and show that the current most stringent constraints on \ac{PBH} abundance can give a measurable \ac{SGWB} in upcoming observing runs of Advanced LIGO. The \ac{SGWB} from \acp{PBH} provides a complementary channel to investigate the existence of subsolar mass \acp{BH}, which is a smoking gun for \acp{PBH}, even if their \ac{GW} signals are too weak to be resolved individually.

\textit{Merger rate of primordial black hole binaries.}--- We give a brief overview of the formation mechanism of the binary \ac{PBH} mergers proposed in Ref.~\cite{Nakamura:1997sm} and revisited by Refs.~\cite{Ioka:1998nz,Sasaki:2016jop} to study the merger rate of \ac{PBH} binaries and the \ac{SGWB} from \ac{PBH} binary merger. 
\acp{PBH} are formed deep in the radiation-dominated epoch and decouple from the background when the average energy density of \acp{PBH} exceeds the background cosmic energy density.
The tidal force from a third \ac{PBH} causes a \ac{PBH} pair to move along elliptical orbits and finally to merge due to the energy loss via gravitational radiation.
Assuming the abundance of \acp{PBH} in dark matter to be $f$ (i.e., $\Omega_{\textrm{PBH}} = f \Omega_\textrm{DM}$), and a fixed \ac{PBH} mass $M_{\rm PBH}$, the probability that the coalescence occurs in the cosmic time interval $(t,t+dt)$ is given by
\begin{align}
	dP_t=
	\begin{cases}
		\frac{3}{58}\left[-\left(\frac{t}{T}\right)^{\frac{3}{8}}+\left(\frac{t}{T}\right)^{\frac{3}{37}}\right]\frac{dt}{t},~~~~~~~~~~~~\textrm{for}~ t< t_c\\
		\frac{3}{58}\left(\frac{t}{T}\right)^{\frac{3}{8}}\left[-1+\left(\frac{t}{t_c}\right)^{-\frac{29}{56}}f^{-\frac{29}{8}}\right]\frac{dt}{t},\textrm{for}~ t\geq t_c,
	\end{cases}
	\label{dpt}
\end{align}
where $T=\frac{3}{170}\frac{c^5{\bar{x}}^4}{ (G M_{\textrm{PBH}})^3 f^4}$ and $t_c = \frac{3}{170}\frac{c^5 {\bar{x}}^4 f^{25/3}}{ (G M_{\textrm{PBH}})^3}$ are constants, $c$ is the speed of light, $G$ is the gravitational constant, and $\bar{x}$ is the physical mean separation of \acp{PBH} at the epoch of matter-radiation equality when redshift $z= z_{\textrm{eq}}$ \cite{Sasaki:2016jop}.

The merger rate of \ac{PBH} binaries is then given by
\begin{align}
	R_{\textrm{PBH}}(z;M_\text{PBH},f)=\frac{3H_0^2}{8\pi G}\frac{f\Omega_{\textrm{DM}}}{M_{\textrm{PBH}}}\frac{dP_t}{dt}.
	\label{merger rate}
\end{align}
Here the redshift $z$ is related to the cosmic time $t$ by $t=t_0-\frac1{H_0}\int_{0}^{z}\frac{dz^\prime}{(1+z^\prime)E(z^\prime)}$, where $t_0$ denotes the age of the Universe and $E(z)\equiv H(z)/H_0=\left[\Omega_{\textrm{r}}(1+z)^4+\Omega_{\textrm{M}}(1+z)^3+\Omega_{\Lambda}\right]^{{1}/{2}}$.
Throughout this work, we use the cosmological parameters derived from the 2015 Planck data set \cite{Ade:2015xua}, i.e., the Hubble constant $H_0=67.8~\textrm{km}~\textrm{s}^{-1}\textrm{Mpc}^{-1}$, the fraction of radiation $\Omega_{r}=9.061\times 10^{-5}$, the fraction of dark matter $\Omega_{\textrm{DM}}=0.270$, the fraction of nonrelativistic matter $\Omega_{M}=0.307$ and the fraction of dark energy $\Omega_\Lambda=1-\Omega_{M}-\Omega_{\textrm{r}}$.
For the \ac{PBH} mass spectrum, a narrow spread distribution \cite{Kawasaki:2012wr,Jedamzik:1999am,Kodama:1982sf,Khlopov:2000js} and an extended distribution \cite{Carr:1975qj,Hawking:1987bn} are both considered by early Universe models.
However, it has been shown that the inflationary scenario does not favor those with a significantly extended \ac{PBH} mass distribution \cite{Carr:2016drx}.
We also find that for a Gaussian \ac{PBH} mass distribution with a narrow width $\Delta M\sim1M_\odot$ the resulting \ac{SGWB} amplitude is negligibly different (less than 1\% between $10-100\rm{Hz}$) from that by assuming a fixed mass distribution.
A later work by Ref.~\cite{Raidal:2017mfl} that generalized our constraining results also confirmed that assuming a log-normal mass distribution with variance $\sigma\sim O(1)$ would not change the order of magnitude of the upper limits on \ac{PBH} abundance. 
Therefore, given the large uncertainties in the \ac{PBH} mass distribution \cite{Carr:2016drx} and aiming to investigate to which extent \ac{SGWB} can constrain the \ac{PBH} abundance, we follow Sasaki \textit{et al.} \cite{Sasaki:2016jop} and use the simplifying assumption that all \acp{PBH} have the same mass.

In contrast to binary \acp{PBH}, the astrophysical \ac{BBH} merger rate $R_\textrm{astro}(z)$, which is described in e.g., Ref.~\cite{TheLIGOScientific:2016wyq}, peaks at $z=1\sim2$, because of the peak in the astrophysical star formation rate \cite{Vangioni:2014axa}.
While for the \ac{PBH} binaries whose mass and local merger rate are consistent with those of GW150914 and GW151226, the merger rate $R_\textrm{PBH}(z)$ keeps rising out to a large redshift (at least $z\sim30$, see Ref.~\cite{Nakamura:2016hna}), due to the fact that \acp{PBH} form in the early Universe, and thus have a larger merger rate at high redshift than astrophysical \acp{BBH}.

The merger rate as a function of redshift, especially at high redshift, can give us important information about the origin of \acp{BBH}, since $R_\textrm{PBH}(z)$ and $R_\textrm{astro}(z)$ behave differently.
Recently it has been proposed that Pre-DECIGO (pre-DECihertz laser Interferometer Gravitational-wave Observatory) can determine the origin of GW150914-like \acp{BBH} by measuring the mass spectrum and the $z$ dependence of the merger rate \cite{Nakamura:2016hna}.
Therefore, future space-based \ac{GW} detectors, such as LISA \cite{Heinzel:2014ets,Bartolo:2016ami}, DECIGO \cite{Kawamura:2006up}, and BBO \cite{Harry:2006fi}, may also be used to study the origin of \acp{BBH}.
The correlation of \ac{GW} events with galaxy catalogs may also distinguish the origin of \acp{BBH} \cite{Raccanelli:2016cud}.

\textit{Stochastic gravitational-wave background energy density spectrum.}---Given the merger rate of \acp{BBH}, one can obtain the \ac{SGWB} energy density spectrum from
\begin{align}
\Omega_{\textrm{GW}}=\frac{\nu}{\rho_c}\frac{d\rho_{\textrm{GW}}}{d\nu},
\end{align}
where $d\rho_{\textrm{GW}}$ is the gravitational radiation energy density in the frequency interval $(\nu,\nu+d\nu)$, and $\rho_c = 3H_0^2 c^2/8\pi G$ is the critical energy density of the Universe \cite{Allen:1997ad}.
For the \ac{SGWB} produced by binary \ac{PBH} mergers, $\Omega_{\textrm{GW}}$ can be expressed as an integral over the redshift, namely
\begin{align}
\Omega_{\textrm{GW}}(\nu;M_\textrm{PBH},f) = &\frac{\nu}{\rho_c H_0} \int_{0}^{z_{\textrm{sup}}} \frac{R_\textrm{PBH}(z;M_\textrm{PBH},f)}{(1+z)E(z)}  \\ \nonumber
& \times \frac{dE_\textrm{GW}}{d\nu_s}(\nu_s) \, dz,
\label{GW spectrum}
\end{align}
where ${dE_\textrm{GW}}/{d\nu_s}(\nu_s)$ is the \ac{GW} energy spectrum of \acp{BBH} coalescence, $\nu_s$ is the frequency in the source frame and is related to the observing frequency $\nu$ through $\nu_s=(1+z)\nu$. The factor $(1+z)$ on the denominator converts the merger rate from source frame to the observer frame.
For this work, we assume an inspiral-merger-ringdown energy spectrum with nonprecessing spin correction \cite{Ajith:2007kx,Ajith:2009bn}.
The upper limit of the integration is given by $z_{\textrm{sup}}=\min(z_{\textrm{max}},\nu_{\textrm{cut}}/\nu-1)$, where $\nu_\textrm{cut}$ is the cutoff frequency given the energy spectrum of the \ac{BBH} and $z_{\textrm{max}}$ is the maximum redshift predicted by the \ac{PBH} model.
Since \acp{PBH} are formed in the early Universe, $z_{\rm max}$ is larger than $\nu_{\textrm{cut}}/\nu-1$ so that $z_{\rm sup}$ never takes the value of $z_{\rm max}$ in the Advanced LIGO sensitive frequency band.

Ref.~\cite{Mandic:2016lcn} investigates the \ac{SGWB} energy density spectrum from \ac{PBH} binaries, compares it to that from astrophysical \acp{BBH} and discusses the detectability for future \ac{GW} detectors.
The \ac{PBH} background was shown to have the same power law spectrum $f^{2/3}$ as that from astrophysical \acp{BBH} in the Advanced LIGO sensitivity band.
Moreover, it has been suggested that the SGWB can be used to investigate the \ac{PBH} abundance. 
Here, we consider the constraints on the \ac{PBH} abundance using the \ac{SGWB} in a different \ac{PBH} binary formation framework by \textcite{Sasaki:2016jop} which produces binaries in the early Universe, as opposed to that used by Ref.~\cite{Mandic:2016lcn} which forms binaries in the late Universe.
Since the \acp{PBH} in the early Universe are distributed more densely, the \textcite{Sasaki:2016jop} framework has a larger merger rate, leading to a stronger \ac{SGWB} amplitude compared with Ref.~\cite{Mandic:2016lcn}.

\textit{Constraining the primordial black hole abundance with the stochastic gravitational-wave background.}---Since the first Advanced LIGO observation run did not find evidence for a \ac{SGWB} signal \cite{TheLIGOScientific:2016dpb}, we can use the nondetection to constrain the maximum \ac{SGWB} energy density spectrum $\Omega_{\textrm{GW}}^{\textrm{max}}(\nu)$, and to further constrain the maximum \ac{PBH} abundance $f_\textrm{max}$ by equating
\begin{equation}
\Omega_{\textrm{GW}}^{\textrm{max}}(\nu) = \Omega_{\textrm{GW}}(\nu;M_\textrm{PBH},f_\textrm{max}),
\end{equation}
thereby giving a upper limit $f_\textrm{max}$ on the \ac{PBH} abundance for different $M_\textrm{PBH}$. 
Taking advantage of the unique observational window from \ac{GW}, the \ac{SGWB} yields a new independent constraint on the properties of \acp{PBH} which we can compare to other methods, such as the lensing of stars and quasars, dynamics of dwarf galaxies, large scale structure and accretion effects on the \ac{CMB} \cite{Carr:2016drx}. 

\begin{figure}[htbp]
	\centering
	\includegraphics[width=\columnwidth]{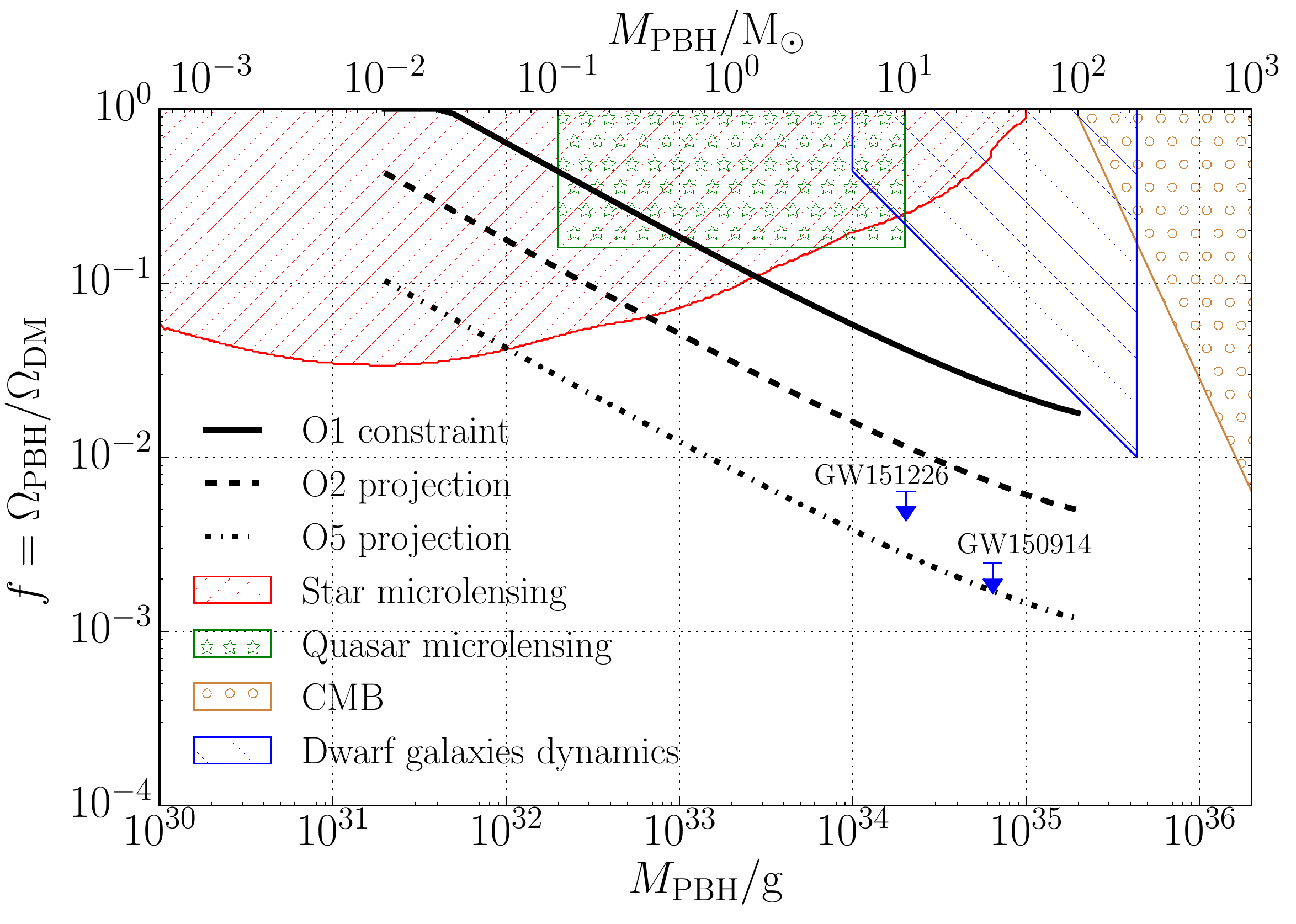}
	\caption{
		The constraints on the \ac{PBH} fraction in dark matter $f_\textrm{max}$ versus the \ac{PBH} mass $M_{\rm PBH}$ from the nondetection of the \ac{SGWB} from Advanced LIGO's O1 and the expected constraints based on the O2 and O5 projected sensitivities.
		These constraints are compared to those from star microlensing \cite{Novati:2013fxa}, quasar microlensing \cite{Mediavilla:2009um}, dynamics of dwarf galaxies \cite{Koushiappas:2017chw} and accretion effects on \ac{CMB} \cite{Ali-Haimoud:2016mbv}.
		The local merger rate for GW150914- and GW151226-like \acp{BBH} can also constrain the \ac{PBH} abundance with corresponding mass. 
	}
	\label{fig:constraint}
\end{figure}

Figure~\ref{fig:constraint} shows the current upper limit in the $f$-$M_{\textrm {PBH}}$ plane from Advanced LIGO's first observation run (O1, 2015–16, black solid), and the expected constraints from the second observation run (O2, 2016–17, black dashed) and the fifth observation run (O5, 2020–22, dot dashed).
For comparison, constraints on $f$ from the EROS-OGLE microlensing of stars (This result is obtained by combining EROS and OGLE detections and achieved tighter constraints by  assuming that a few positive detections from OGLE are explained by self-lensing.) \cite{Novati:2013fxa} , microlensing of quasars \cite{Mediavilla:2009um}, dynamics of dwarf galaxies \cite{Koushiappas:2017chw}, and accretion effect on \ac{CMB} \cite{Ali-Haimoud:2016mbv} are also plotted. 

In addition, the inferred local merger rates associated with the GW150914- and GW151226 like \acp{BBH} can also constrain the abundance of \acp{PBH}. Since we adopt a delta PBH mass distribution following Sasaki \textit{et al.} \cite{Sasaki:2016jop} given the large theoretical uncertainty, for consistency, we also consider the LIGO's local merger rate estimated by assuming all the black holes have the same mass as detected rather than an extended distribution. By imposing the condition that $R_{\textrm{PBH}}(z=0;M_\text{PBH},f_\textrm{max})$ cannot exceed the maximum of the estimated local merger rate, an upper limit on the \ac{PBH} abundance $f_\textrm{max}$ can be given with corresponding $M_\textrm{PBH}$, as shown in \Cref{fig:constraint}.

We see that up to now microlensing gives the tightest constraints in the mass range $10^{-3}M_\odot\lesssim M_\text{PBH}\lesssim 1 M_\odot$. The new upper limit given by \ac{SGWB} from Advanced LIGO's O1 gives the best constraint on the \ac{PBH} abundance in the mass range $1 M_\odot \lesssim M_\text{PBH}\lesssim 100 M_\odot$ (
\Ac{PBH} binaries with masses higher than $\sim O(10^2)M_\odot$ would have lower cut-off frequency, thus would evade the frequency band of Advanced LIGO),
pushing the previous microlensing and dwarf galaxy dynamics constraint tighter by 1 order of magnitude.
Future observing runs of Advanced LIGO are expected to improve the constraint $f_\textrm{max}$ further to $O(10^{-3})$. 

\begin{figure}[htbp]
	\centering
	\includegraphics[width=\columnwidth]{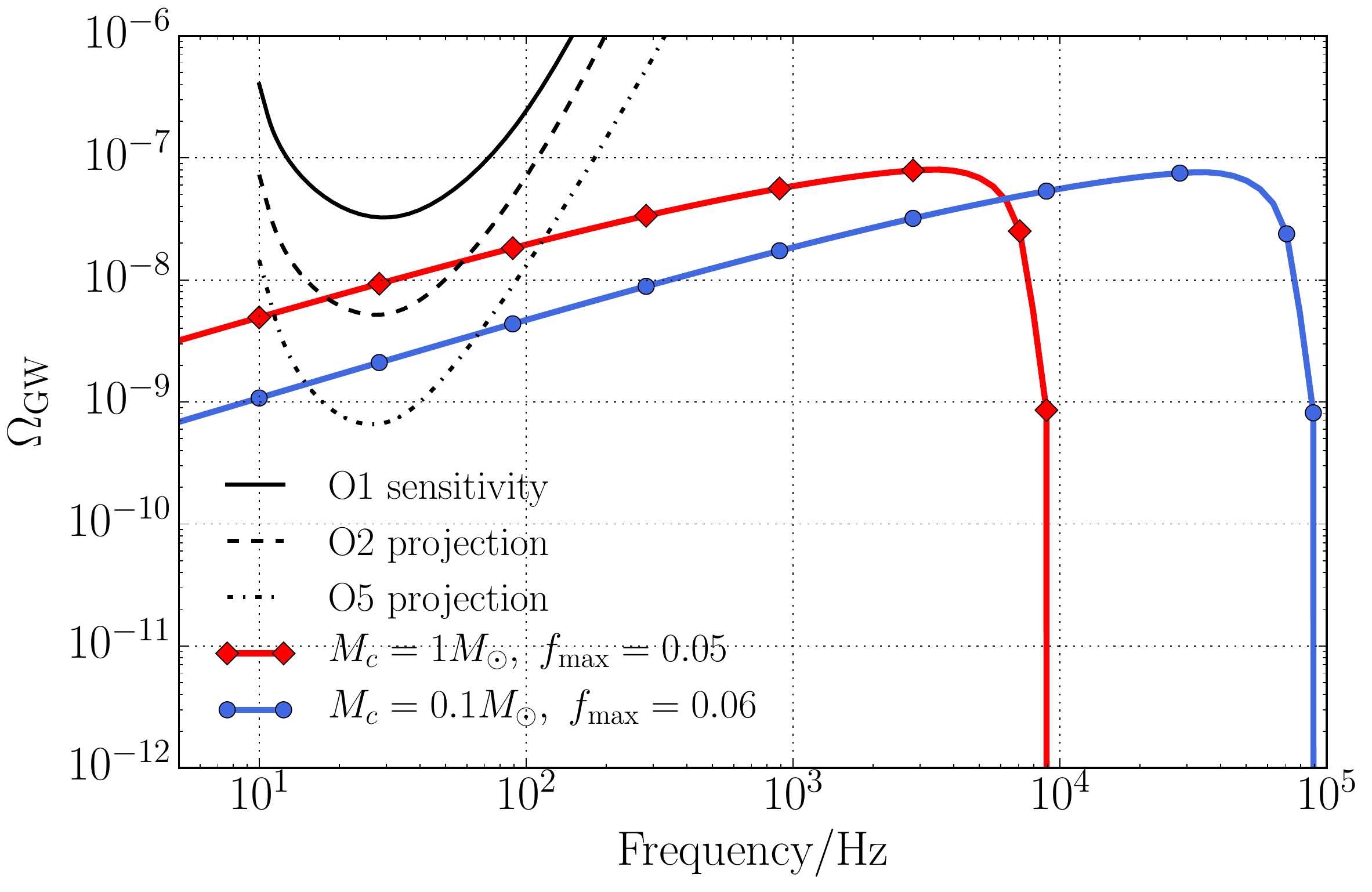}
	\caption{
		The \ac{SGWB} spectra from subsolar mass binary \ac{PBH} mergers at the current best constraints from stellar microlensing.
		Nondetection in Advanced LIGO's O2 can further constrain the existence of subsolar mass \acp{PBH}.
	}
	\label{fig:ogwfm}
\end{figure}

Conversely, we can compare the \ac{SGWB} spectra from the current constraints to the expected Advanced LIGO sensitivities. 
\Cref{fig:ogwfm} shows the \ac{SGWB} spectra from binary \ac{PBH} mergers for different chirp masses using the current most stringent constraints of the \ac{PBH} abundance.
Here, the chirp mass is defined by $M_c=(m_1m_2)^{3/5}/(m_1+m_2)^{1/5}$, where $m_1$ and $m_2$ are the component mass of \acp{BBH}. Thus, $M_c = M_\textrm{PBH}/2^{1/5}$ for \acp{PBH} of a fixed mass.
In \Cref{fig:ogwfm}, the black curves denote the $1\sigma$ sensitivity of the LIGO-Virgo network expected for two first observing runs O1 (black solid) and O2 (black dashed), and at the design sensitivity in O5 (black dot dashed) \cite{TheLIGOScientific:2016wyq,Thrane:2013oya}. The sensitivity curve is calculated in the context of the cross correlation statistic method \cite{Allen:1997ad} with one year of integration, and the coincident duty cycle is $30\%$ for O1 and $50\%$ for O2 and O5. If a model-predicting spectrum intersects a black curve, then it has an expected signal-to-noise ratio greater or equal than 1.

From \Cref{fig:ogwfm} we can see that the \ac{SGWB} generated by subsolar mass \acp{PBH} has the opportunity to be detected by upcoming Advanced LIGO observing runs.
Therefore, the SGWB provides a possible way to explore the existence of subsolar mass \acp{PBH}, which would be a smoking gun for the existence of \acp{PBH} since subsolar mass \acp{BH} are not expected to be of a stellar origin. 
However, a decisive evidence for \acp{PBH} would need the detection of \ac{SGWB} at high frequency, which is beyond the scope of Advanced LIGO.
Nevertheless, \ac{SGWB} provides a complementary channel to investigate the properties of subsolar mass \acp{BH}, even if their \ac{GW} signal is too weak to be individually resolved.

\textit{Discussion.}---In this work, we place a novel constraint on the \ac{PBH} abundance in dark matter in the mass range $0.01M_\odot -100M_\odot$ using the current nondetection of \ac{SGWB} from Advanced LIGO's first observation run.
As a new observational window, the constraint from \ac{SGWB} is better than other methods such as microlensing and dwarf galaxy dynamics by 1 order of magnitude in the mass range $1 M_\odot \lesssim M_\text{PBH}\lesssim 100 M_\odot$.
Finally, we also find that the current most stringent constraints on the abundance of subsolar mass \acp{PBH} can give a measurable \ac{SGWB} by future Advanced LIGO observing runs.

The coalescence of a pair of \acp{PBH} produces a BH of higher mass, and this evolution of the mass distribution has an effect on the \ac{SGWB} spectrum.
Nevertheless, at the matter-radiation equality epoch $z_\mathrm{eq}$, only a pair of \acp{PBH} that satisfies $x<f^{1/3} \bar x$ can form a binary \cite{Sasaki:2016jop}, where $x$ is the physical separation between two neighboring \acp{PBH}.
This means that the fraction of \acp{PBH} that can form binaries in the total \ac{PBH} population is $x^3_{\mathrm{max}}/{\bar x}^3 \simeq f$.
Thus, the fraction of subsequent more-massive \ac{PBH} binaries in the original population of \ac{PBH} binaries is also given by $f$.
From \cref{fig:constraint}, the typical value of $f_\textrm{max}$ is of the order of $O(0.01)$.
However, the mass doubling effect will only contribute an extra factor of $2^{5/3}\sim3$ 
to the \ac{GW} energy density spectrum ($dE/d\nu\propto M_c^{5/3}$).
Therefore, we expect that the evolution of the mass distribution has a negligible effect on the \ac{SGWB} in this work's scenario.
However, we also notice that \acp{PBH} may be clustered in the late Universe, boosting the formation rate of more-massive \ac{PBH} binaries.
The effect of such clustering is beyond the scope of this Letter and is left for a future work.

Another consideration is that \ac{PBH} binaries may be formed with highly eccentric orbits, and these binaries could preserve the eccentricity until merger if they coalesce on timescales within years or less \cite{Cholis:2016kqi,Clesse:2016ajp}.
In this work, the contribution of \ac{PBH} mergers to the \ac{SGWB} spectrum in the Advanced LIGO sensitive frequency band comes from the redshift $z<\nu_{\textrm{cut}}/\nu-1$.
Compared with the binary formation epoch, which is earlier than the matter-radiation epoch $z_\mathrm{eq}$, the \ac{PBH} binaries are expected to have enough time to circularize the orbits.
Therefore, we assume that the effects of eccentricity are negligible when calculating the \ac{SGWB} in this work.
However, when considering a lower frequency band, one should include the influence of the gravitational-wave emission from eccentric binaries (see, e.g., Ref.~\cite{Yunes:2009yz,PhysRevD.90.084016,Mikoczi:2015ewa,Tanay:2016zog,Mishra:2015bqa,PhysRevD.95.024038,Chen:2016zyo}) on the \ac{SGWB}.

Finally, our results depend on the merger rate of Sasaki \textit{et al.} \cite{Sasaki:2016jop}, which in turn assumes that binaries formed at early times and not merged yet survive until $z=0$. 
The \ac{PBH} binary formation and evolution are still under active investigation (see, \textit{e.g.} Ref.~\cite{Hayasaki:2009ug,Eroshenko:2016hmn,Ali-Haimoud:2017rtz}).
Future \ac{GW} measurements will further shed light on the \ac{PBH} scenario.

Recently, three more \ac{GW} events from \ac{BBH} merger, GW170104, GW170608 and GW170814, were announced during the second Advanced LIGO-Virgo observation run \cite{Abbott:2017vtc,Abbott:2017gyy,Abbott:2017oio}.
In the absence of a publicly available event rate statement from each event alone and \ac{SGWB} results for the second observation run, we leave this analysis for future work.

\vspace{0.1cm}

\begin{acknowledgements}
S.W. is partially supported by the funding from the Research Grants Council of Hong Kong (Project No. 14301214 and ECS No. 2191110). Y.-F.W and T.G.F.L were partially supported by a grant from the Research Grants Council of Hong Kong (Project No. CUHK 14310816) and by the Direct Grant for Research from the Research Committee of the Chinese University of Hong Kong. Q.-G.H. is supported by grants from NSFC (Grants No. 11335012, No. 11575271, and No. 11690021), and the Top-Notch Young Talents Program of China, and is partly supported by the Key Research Program of Frontier Sciences, CAS. Q.-G.H. also acknowledges the hospitality of HKUST during the last stage of this project.
\end{acknowledgements}

\bibliography{sgwb_pbh}
\end{document}